\documentclass[sigconf]{acmart}
\usepackage[table,xcdraw]{xcolor}
\usepackage{siunitx}
\usepackage{mdframed}

\setcopyright{none}
\acmConference[AST '26]{7th ACM/IEEE International Conference on Automation of Software Test (AST 2026)}{April 13--14, 2026}{Rio de Janeiro, Brazil}

\newmdenv[
  leftline=true,
  topline=false,
  bottomline=false,
  rightline=false,
  linecolor=gray,
  linewidth=2pt,
  skipabove=\baselineskip,
  skipbelow=\baselineskip,
  innerleftmargin=10pt,
  innerrightmargin=10pt,
  innertopmargin=5pt,
  innerbottommargin=5pt
]{customquote}

\makeatother

\begin{document}

\title{Understanding on the Edge: LLM-generated Boundary Test Explanations}

\author{Sabinakhon Akbarova}
\orcid{0000-0002-6518-8193}
\email{sabina.akbarova@chalmers.se}
\affiliation{%
  \institution{Chalmers University of Technology and University of Gothenburg}
  \city{Gothenburg}
  \country{Sweden}
}

\author{Felix Dobslaw}
\orcid{0000-0001-9372-3416}
\email{felix.dobslaw@miun.se}
\affiliation{%
  \institution{Mid Sweden University}
  \city{Östersund}
  \country{Sweden}
}

\author{Robert Feldt}
\orcid{0000-0002-5179-4205}
\email{robert.feldt@chalmers.se}
\affiliation{%
  \institution{Chalmers University of Technology and University of Gothenburg}
  \city{Gothenburg}
  \country{Sweden}
}

\renewcommand{\shortauthors}{Akbarova, Dobslaw, and Feldt}


\begin{abstract}
Boundary value analysis and testing (BVT) is fundamental in software quality assurance because faults tend to cluster at input extremes, yet testers often struggle to understand and justify why certain input-output pairs represent meaningful behavioral boundaries. Large Language Models (LLMs) could help by producing natural-language rationales, but their value for BVT has not been empirically assessed. We therefore conducted an exploratory study on LLM-generated boundary explanations: in a survey, twenty-seven software professionals rated GPT-4.1 explanations for twenty boundary pairs on clarity, correctness, completeness and perceived usefulness, and six of them elaborated in follow-up interviews. Overall, 63.5\% of all ratings were positive (4–5 on a five-point Likert scale) compared to 17\% negative (1–2), indicating general agreement but also variability in perceptions. Participants favored explanations that followed a clear structure, cited authoritative sources, and adapted their depth to the reader’s expertise; they also stressed the need for actionable examples to support debugging and documentation. From these insights, we distilled a seven-item requirement checklist that defines concrete design criteria for future LLM-based boundary explanation tools. The results suggest that, with further refinement, LLM-based tools can support testing workflows by making boundary explanations more actionable and trustworthy.

\end{abstract}

\maketitle
\noindent
\textbf{Author’s Accepted Manuscript.}
This is the author’s version of a paper accepted for publication in
\emph{Automation of Software Test (AST 2026)}.
The final version will be available via the ACM Digital Library.

\section{Introduction}
Boundary value analysis and testing (BVT) is a textbook technique; taught early, used often, and widely seen as essential for effective test suites. It rests on a simple insight: software often breaks at the ``edges'' between \textit{equivalence partitions}, where one behavioral region meets another~\cite{reid1997empirical}. Yet despite its ubiquity, BVT has long lacked a precise formal definition, limiting its theoretical foundations and the development of systematic tools support \cite{hierons2006avoiding, bhat2015equivalence}. Practitioners have typically relied on informal rules, intuition, or manual reasoning to identify meaningful boundaries.

In practice, the applicability of BVT varies considerably with the structure of the input domain. For functions with simple, well-defined input types, partitions and boundary values are often straightforward to identify. However, BVT becomes more challenging for software with large input spaces, complex or highly structured data types, or strong dependencies between inputs, where specifications are often incomplete or expressed only informally. In such cases, identifying meaningful partitions and boundaries requires human interpretation, which limits both reliability and automation \cite{dobslaw2020boundary}.

Recent work has begun to change this based on a quantification of \textit{boundariness}, the degree to which nearby inputs produce different outputs~\cite{feldt2019towards}. This enabled new approaches for augmenting human boundary exploration~\cite{dobslaw2020boundary} and the first automated tools capable of discovering boundaries without human guidance~\cite{dobslaw2023automated}. But identification is only part of the problem: since boundaries often lack obvious correctness cues, testers need explanations to understand \textit{why} a given input-output pair lies ``on the edge''. Explanations are therefore a prerequisite for effective boundary identification, offering both interpretability and a means to validate automated discoveries.

Large Language Models (LLMs) have transformed software engineering tasks --- from test generation to code review --- by combining pattern recognition with approximate reasoning~\cite{wang2024software}. They have been proposed as drivers for autonomous testing agents~\cite{feldt2023towards} and achieved state-of-the-art results in complex domains like mobile app testing through multi-agent collaboration~\cite{yoon2024intent}. Could they enhance automated boundary testing by using their code and language understanding to recognize boundary patterns or generate plausible candidates?

We argue that explanation is a necessary foundation for applying LLMs in boundary testing. Even technically successful boundary identification tools will fail in practice if they cannot justify their suggestions to skeptical developers and testers --- much like auto-generated test cases that lose value when they are not human-readable~\cite{fraser2015does}. 
Moreover, our vision centers on human-AI collaboration~\cite{feldt2023towards}, where testers, not black-box automation, make final decisions about test suite. Explanation also provides a tractable way to assess whether an LLM conceptually understands boundaries before attempting autonomous identification.

This paper presents the first exploratory systematic investigation into whether LLMs can provide meaningful boundary explanations. As prior work has not examined this topic, no standard baselines exist. We therefore employ a human-centric evaluation to assess their perceived quality without assuming a gold standard, focusing on whether LLMs can clearly articulate why a given input–output pair represents a behavioral boundary. This study marks the first step in a two-stage research program, where explanation serves as a low-stakes yet rigorous entry point toward using LLMs for boundary \textit{identification} and assessing their conceptual understanding of boundaries.

To this end, we conducted an empirical study with software professionals, who evaluated LLM-generated explanations for manually selected and automatically discovered boundary candidates, followed up by semi-structured interviews with six participants. To capture perspectives from both research and practice, our sample included both software researchers and industry practitioners.
Specifically, we investigate the following research questions:
\begin{itemize}
    \item \textbf{RQ1}: Can LLMs be used to explain software boundaries to testers? The aspects we look for are (a) clarity, (b) correctness, (c) completeness, (d) and perceived usefulness.
    \item \textbf{RQ2}: What makes for good boundary explanations, and how can they be used?
\end{itemize}

To the best of our knowledge, this is the first study addressing the question of the explainability of boundaries in software. The question order is intentional since we initially do not know what criteria a boundary explanation ideally fulfills. To broadly capture expert assessment, we needed criteria and hypothesized, in support of the other literature on explainability \cite{liao2021human, ji2023survey, howcroft2020twenty}, that clarity, correctness, completeness, and usefulness were all important and suitable. Survey and interview responses would offer solicited feedback on the perceived explainability and the suitability of the criteria for that assessment alike.

The main contributions of this paper are threefold. First, we present the first empirical study on LLM-generated explanations for software boundary cases, based on a survey and follow-up interviews with software professionals from both research and industry. Second, we provide a detailed evaluation of explanation quality across four key dimensions --- clarity, correctness, completeness, and usefulness, based on the respondents' feedback. Third, we identify practical requirements and improvement opportunities for LLM-based boundary explanation tools, manifested in a 7-item requirement list for future research and tool development. Together, these contributions offer new insights into how LLMs can enhance the explainability and adoption of automated boundary testing in practice and beyond.

The rest of this paper is structured as follows. Section~\ref{bgrw} outlines background on boundary value testing, LLMs in software testing, and related work on test selection and LLM-based tools. Section~\ref{method} details our survey design, data collection, and analysis. Section~\ref{results} presents findings on how participants perceived the LLM-generated explanations. Section~\ref{discussion} interprets the results and reflects on limitations and validity. Section~\ref{conclusions} concludes the paper and suggests directions for future work.

\section{Background and Related Work}
\label{bgrw}

BVT has been widely adopted since the 1990s, yet early studies noted the lack of a precise formal model, which hampered rigorous tool support and theory development~\cite{reid1997empirical,hierons2006avoiding,bhat2015equivalence}. More recently, \emph{boundariness} metrics have enabled search‐based and derivative‐based techniques that automatically uncover edge cases without domain expertise~\cite{feldt2019towards,dobslaw2020boundary,dobslaw2023automated}. These algorithms reliably surface candidate boundaries, but still rely on humans to interpret the behavioural significance of the results.

Parallel advances in LLMs have pushed the frontier of automated test generation, autonomous testing agents, and code  review~\cite{wang2024software,feldt2023towards,yoon2024intent}.  However, existing work treats LLMs primarily as generators; their potential to \emph{explain} why a boundary matters remains unexplored. Prior empirical evidence shows that developers often discard auto-generated tests lacking transparent justification~\cite{fraser2015does} and that trust can shift once AI authorship is revealed~\cite{lim2024effect}.

Further, advancements in code summarisation leverage LLMs to generate human-readable rationales for source code and have demonstrated that LLMs like CodeLlama and GPT-4 outperform previous models in generating accurate and context-aware summaries, particularly at the statement level, thereby enhancing code comprehension and maintenance \cite{zhu2024effectiveness}. Beyond documentation, researchers have explored explanations for static-analysis warnings and code review comments, showing that clarity and context reduce false-positive fatigue \cite{johnson2013don}. Still, developers routinely discard or distrust auto-generated outputs that lack adequate justification or appear inconsistent with domain knowledge \cite{fraser2015does}.

To capture professionals’ assessment, Human-Computer Interaction (HCI) studies emphasise soundness, sufficiency, and clarity as core qualities of effective explanations~\cite{liao2021human}, while correctness is central to evaluating LLM outputs~\cite{ji2023survey}. Howcroft et al.~\cite{howcroft2020twenty} further identify clarity, correctness, and usefulness as common NLG criteria, with completeness often covered by information content or understandability. The four empirical criteria used in this study --- clarity, correctness, completeness, and usefulness --- thus reflect both conceptual relevance and established practice, extending explainability research into the novel domain of BVT.

Taken together, these threads suggest a pressing need to study whether LLMs can supply concise, correct, and useful explanations for automatically discovered boundaries --- an open question we tackle in this paper.

\section{Method}
\label{method}

\subsection{Study Design and Rationale}

We employed a \emph{sequential mixed-methods} design consisting of (i) an online survey and (ii) follow-up semi-structured interviews. In the survey, we did not reveal that the explanations were generated by an LLM, as such framing has in other contexts been shown to produce significantly more negative outcomes~\cite{lim2024effect}. The survey captured broad quantitative judgments of automatically generated boundary-value explanations, while the interviews added depth by probing participants’ reasoning once they learned about the LLM source. This sequence avoided anchoring effects, allowed the interview guide to be refined from survey free-text, and triangulated findings to strengthen construct validity. For transparency and reproducibility, all materials --- including prompts, survey and interview questions, responses, and scripts --- are available in a replication package\footnote{\href{https://zenodo.org/records/17434668?preview=1&token=eyJhbGciOiJIUzUxMiJ9.eyJpZCI6IjM4MWM3NmU1LWNhYTUtNDI5MS1iMTVlLTI0NjI2MmE3ODM5OCIsImRhdGEiOnt9LCJyYW5kb20iOiI0YWVhOGE2YzcxMGYyMGY0MmQxNWVkNTQ2ZTliYzQyMCJ9.ani5_hWuM2zGw5-zWhTyKbSWvudUlOZnbop72-PIBbGTH3aagaXaosAU6xt-qui0hnfIqBI5Y9Pja6KHe6zu4w}{zenodo preview link}}.

\subsection{Context and Materials}
\label{sec:context}

\paragraph{Objects under test}

We selected four functions under test (FUTs) that varied in input type, number of parameters, and domain logic, balancing simplicity and the ability to quickly comprehend them. We presented each FUT by its pseudocode function signature, accompanied by a brief description and an example boundary pair.

\begin{itemize}
\item \textbf{Email} --- checks whether a string is a valid email address.
\item \textbf{Bytecount (BC)} --- converts an integer byte count to a human-readable format using SI prefixes.
\item \textbf{Date} --- creates a Gregorian calendar date from a \texttt{(year, month, day)} tuple.
\item \textbf{BMI} --- calculates adult Body Mass Index and categorises weight status.
\end{itemize}

All programs were treated strictly as \emph{black boxes}. The implementation language was not disclosed, and participants were not exposed to source code, syntax, or language-specific constructs, as these were judged not applicable to the task.

\paragraph{Boundary pair generation}
Each FUT was represented by five \emph{boundary pairs}. For the \textbf{Email} FUT, pairs were generated using a ChatGPT-assisted prompt; for the remaining FUTs, boundary pairs were obtained from the open-source datasets provided by the search-based boundary explorer \textsc{SETBVE}~\cite{akbarova2025setbve}, which employed Julia implementations of these FUTs. To ensure diversity, we first selected the top 100 pairs per FUT based on boundary pair quality, measured by program derivative~\cite{feldt2019towards}. We then applied a greedy max-min strategy to extract the 10 most diverse pairs, prioritising output variation. From these, five were chosen using the Valid (V) and Erroneous (E) classification by Dobslaw et al.~\cite{dobslaw2023automated}, ensuring coverage of the VV, VE, and EE groups where possible. We also favoured shorter inputs to support readability and comprehension. All three authors reviewed the pairs and confirmed that they satisfied domain semantics and exhibited diversity.

\paragraph{Explanation generation}
Explanations for each boundary pair were generated in May 2025 using OpenAI’s API with the GPT-4.1 model. 
We used a temperature setting of 0.2, informed by findings from Oli et al.~\cite{oli2023behavior}, who showed that lower temperatures (0.0 -- 0.5) yield more correct, specific, and concise outputs in generation of code explanations. To validate the model choice, we compared GPT-4.0 and GPT-4.1 outputs using cosine similarity across multiple runs; the average similarity of $0.84 \pm 0.09$ across all FUTs indicated highly comparable outputs. Given this and GPT-4.1’s improved performance at a lower cost \cite{openai2025gpt41}, we proceeded with GPT-4.1. 

Each request included a system prompt framing the model as an expert software tester and a user prompt with the function signature, a one-sentence FUT description, and an input-output pair. The prompt also defined three core requirements and five stylistic guidelines, such as avoiding source code and speculation. We limited explanations to 150 characters to reduce cognitive load, as participants evaluated 20 items across four criteria. This constraint still allowed for meaningful explanations while keeping the task manageable.

\subsection{Participants and Recruitment}

We distributed survey invitations via email and LinkedIn to the authors’ professional contacts, who were encouraged to share the link internally (convenience sampling); the study was not publicly advertised. Any software professional involved in testing could participate. We collected each respondent’s role, years of experience, and familiarity with boundary value analysis. The survey received 27 responses: 13 from researchers and 14 from practitioners. Six volunteers (three researchers, three practitioners) took part in follow-up interviews --- a number deemed sufficient for an exploratory study. Under institutional guidelines, the study was exempt from formal ethics review, as no personal data were required. Participation was anonymous unless an email was provided for interview scheduling; these addresses were deleted after analysis, leaving the dataset anonymised.

\subsection{Survey}
\label{sec:survey}

The survey was conducted using \emph{Netigate} under a university license. The average completion time was 65 minutes.\footnote{Excludes responses submitted more than a day after starting, as these likely involved participants pausing and returning later.}

The survey began with a brief explanation of boundary value analysis and a guiding example, followed by a description of the task: to assess boundary explanations using four criteria across several FUTs. Participants then provided demographic information and answered an open-ended question about what they believe makes for a good boundary explanation.

\begin{figure}
    \centering
    \includegraphics[width=\linewidth]{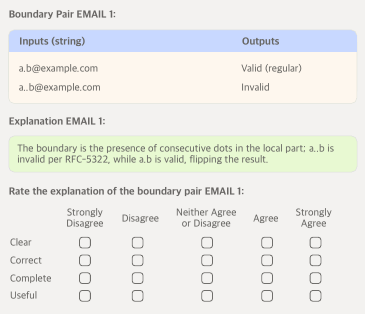}
    \caption{Example of a boundary pair shown to participants with an explanation. The LLM origin of explanations was revealed only during interviews.}
    \label{fig:boundaryexample}
\end{figure}

Each of the four FUTs appeared in a fixed order (Email $\rightarrow$ BC $\rightarrow$ Date $\rightarrow$ BMI) and was introduced by a one-sentence FUT description and a pseudocode method signature. For every boundary pair, participants viewed (see Figure~\ref{fig:boundaryexample}):

\begin{enumerate}
  \item a boundary pair’s inputs and outputs,
  \item the corresponding explanation, and
  \item a four-item five-point Likert matrix rating the explanation’s clarity, correctness, completeness, and usefulness.
\end{enumerate}

Each criterion was accompanied by a tooltip definition, following best practices~\cite{howcroft2020twenty}. The four criteria were defined as follows: \emph{clarity} --- the explanation is easy to read and grasp; \emph{correctness} --- it is factually accurate given the assumptions; \emph{completeness} --- it covers all key aspects given the assumptions; and \emph{perceived usefulness} --- it adds value. All Likert items were mandatory. After each FUT section, participants could provide criterion-specific feedback in free-text fields. In the final section, they were asked whether they would consider using tools based on such explanations, whether they consented to be contacted for interviews, and what single change would make the explanations more useful to them.

\subsection{Interview Procedure}

Interview invitations were sent to the survey respondents who provided contact details; all agreed to participate. Each 30-minute video session followed a semi-structured guide informed by survey comments. The guide included seven main questions with optional probes, starting with the four evaluation criteria, followed by reactions to learning the explanations were LLM-generated, potential use cases, and factors influencing trust or rejection of such tools. Interviews were audio-recorded with consent, transcribed automatically, and manually checked and summarised.

\subsection{Data Analysis}

\paragraph{Quantitative (survey)}  
We computed descriptive statistics on participants’ backgrounds (mean, median, standard deviation), and aggregated explanation ratings per criterion and per FUT.

\paragraph{Qualitative (survey comments \& interviews)}

All survey free-text responses were reviewed by all authors, who jointly developed the follow-up interview questions. Interview transcripts were analysed using reflexive thematic analysis~\cite{braun2006using}. One author performed the initial coding, which was then reviewed and refined collaboratively.

\section{Results}
\label{results}
Results are organised by research question. The first --- whether LLMs can explain software boundaries --- is addressed primarily through quantitative survey data, with free-text comments providing context. The second --- what makes a good boundary explanation --- draws on qualitative data from survey responses and interview analysis.

The survey received 27 responses (21 complete, 6 partial), with an almost even split between software practitioners (14) and researchers (13) and a median experience of ten years. Participants represented diverse domains (multiple selections allowed): AI\slash ML (18), enterprise/backend (14), web\slash cloud (9), embedded systems (6), mobile (5), and games (3). Regarding familiarity with boundary value analysis, 5 participants were \textit{not familiar}, 12 had \textit{some familiarity}, 8 were \textit{quite familiar}, and 2 identified as \textit{experts}.

The interviews were conducted with three practitioners and three researchers who participated in the survey and signalled their interest.

\subsection{RQ1: LLMs and Software Boundary Explanations}
\label{a:rq1}

In the survey, we asked participants to evaluate explanations of five boundary pairs per FUT using a five-point Likert scale across four categories. After completing the evaluations for a given FUT, participants provided free-text suggestions for improvement in each of the four categories. The category-specific summaries of participants’ feedback are presented further in this section. All but two respondents indicated they would consider using automated test generation tools that provide such explanations --- two responded \textit{no}, ten responded \textit{yes}, and the remaining selected \textit{maybe} after completing the survey.

Table \ref{tab:FUT_category_stats} contains the descriptive statistics for all categories over all FUTs. Participants generally agreed with the example explanations: on the five-point Likert scale, all categories averaged above 3.5, with only correctness and usefulness for the Email FUT reaching 4. This indicates overall positive but improvable ratings. In total, 63.5\% of ratings were 4 or 5 (agree or strongly agree), 19.5\% were neutral, and 17\% were 1 or 2 (disagree or strongly disagree).

\begin{table*}[ht]
\centering
\caption{Descriptive Statistics (Mean $\pm$ Std) per FUT and Category}
\begin{tabular}{l *{4}{S[table-format=1.2(2)]}}
\toprule
\textbf{FUT} 
  & \textbf{Clarity} 
  & \textbf{Correctness} 
  & \textbf{Completeness} 
  & \textbf{Usefulness} \\
\midrule
Email      & 3.8 \pm 1.2 & 4.0 \pm 1.1 & 3.7 \pm 1.2 & 4.0 \pm 1.0 \\
Bytecount  & 3.7 \pm 1.1 & 3.8 \pm 1.1 & 3.7 \pm 1.1 & 3.6 \pm 1.3 \\
Date       & 3.6 \pm 1.1 & 3.6 \pm 1.3 & 3.6 \pm 1.1 & 3.6 \pm 1.2 \\
BMI        & 3.8 \pm 1.1 & 3.7 \pm 1.1 & 3.7 \pm 1.1 & 3.5 \pm 1.3 \\
\midrule
\rowcolor{gray!25}
\textbf{Total} & 3.7 \pm 1.1 & 3.8 \pm 1.1 & 3.7 \pm 1.1 & 3.7 \pm 1.2 \\
\bottomrule
\end{tabular}
\label{tab:FUT_category_stats}
\end{table*}

\begin{figure}[H]
    \centering
    \includegraphics[width=\linewidth]{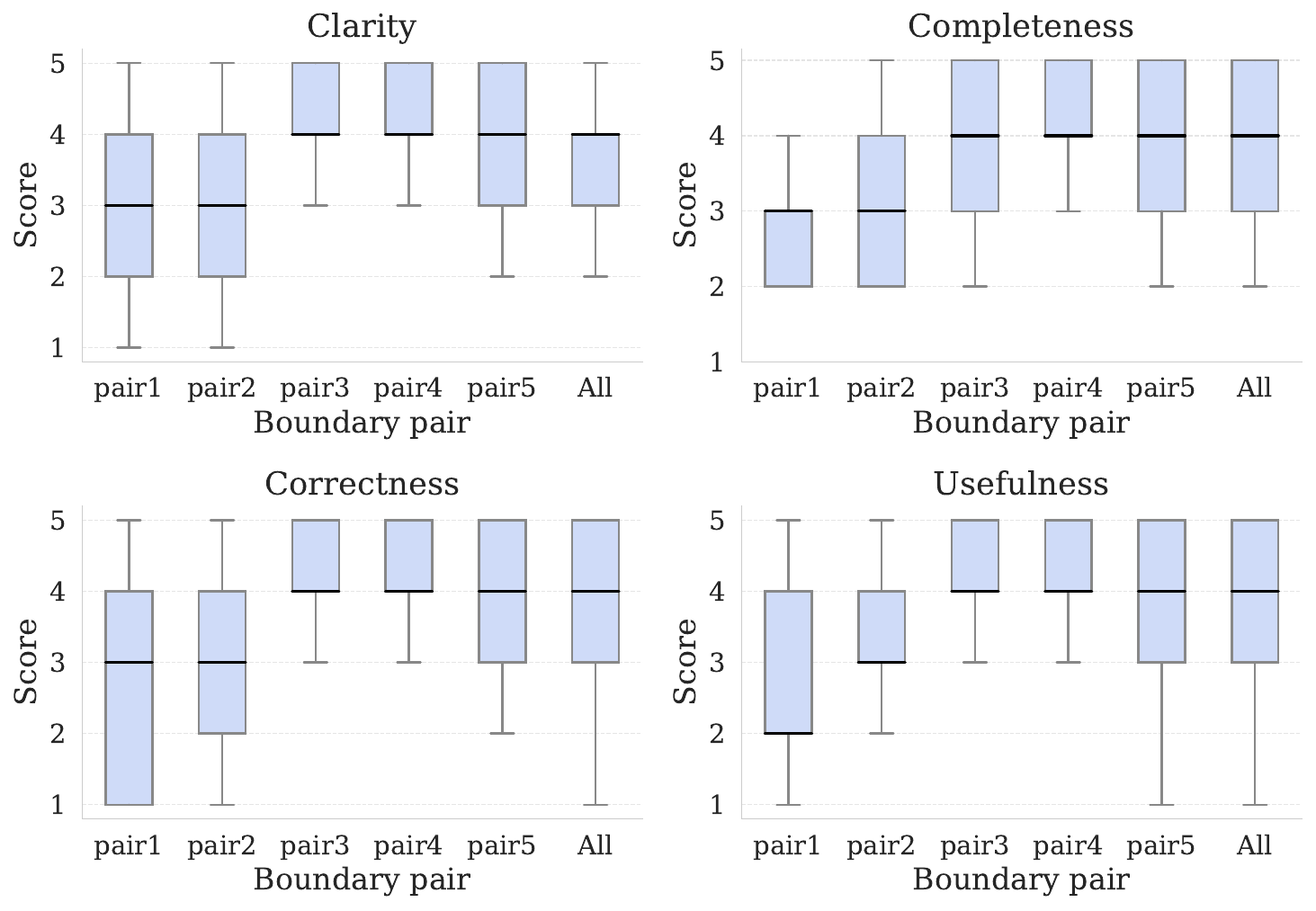}
    \caption{Score distribution for the \textbf{Date} FUT, shown per boundary pair and for all pairs combined. The lower overall ratings --- particularly for correctness --- are due to a hallucination in the explanation of the first boundary pair.}
    \label{fig:dateboxplots}
\end{figure}

Figure \ref{fig:dateboxplots} presents the distribution of evaluation scores for the Date FUT, showing results separately for each boundary pair as well as combined scores across all pairs. We specifically highlight the Date FUT because the first boundary pair exhibits a notable deviation from the overall evaluations presented in Table \ref{tab:FUT_category_stats} (for detailed per-pair distributions of other FUTs, refer to our replication package).

For the Date FUT, clarity, completeness, and correctness received median scores of 3 or higher across all explanations. However, the first pair scored notably low for usefulness (median = 2, “Disagree”) and relatively low for correctness (median = 3, with most ratings between 1–4). This suggests participants largely disagreed that the explanation was accurate or useful. The problematic explanation was for boundary pair of input1: (year=199, month=8, day=5), input2: (year=200, month=8, day=5), with outputs ``0199-08-05” and ``0200-08-05”. The auto-generated explanation stated: \textit{“Crossing year 199 to 200 changes the output’s year field from 3 to 4 digits, assuming all years are valid and zero-padded.”} In reality, both outputs already have four digits, revealing a clear hallucination by the LLM that likely caused the low ratings. No comparable hallucinations were found for other FUTs, and participants offered constructive suggestions for improving explanations across all four evaluation criteria.

\subsubsection{Clarity}

Participants generally found the auto-generated explanations clear and understandable but noted recurring clarity issues across FUTs. The main problems involved undefined technical terms, ambiguous phrasing, and missing contextual details. 

For the Email and Bytecount FUTs, explanations sometimes relied on assumed prior knowledge or unclear notation --- particularly in error messages such as \texttt{BoundsError("kMGTPE”, N)}. Participants recommended simplifying language, defining terms, and explicitly describing thresholds and boundary values.
\begin{customquote}
\textit{“The explanations lack some clarity because to fully understand them, they require prior knowledge of specific terms which are not explained.”}
\end{customquote}

In the Date and BMI FUTs, clarity issues arose from incomplete phrasing and unrealistic examples. Participants noted missing specification details and unreasonable value ranges in some boundary pairs (e.g., a weight of 0 kg in BMI). Several suggested structuring explanations to first describe valid behaviour before outlining the boundary change.

\begin{customquote}
\textit{“The explanation would be clearer and more natural if it describes the valid output first, then how it switches to invalid.”}
\end{customquote}

\subsubsection{Correctness}
Participants generally found the boundary explanations correct but pointed out several issues that reduced their confidence, including  omissions and ambiguous phrasing. They emphasised that correctness depends not only on factual accuracy but also on sufficient context and clear boundary definitions.

For the Email and Bytecount FUTs, participants noted instances of incomplete or unclear explanations, which reduced their confidence in correctness. In the Email FUT, one explanation was seen as incomplete because it failed to clearly define the boundary involving quotation marks around spaces, and the absence of explicit references or context (e.g., relevant RFC standards) further increased uncertainty in correctness. The Bytecount FUT explanations were generally accurate, though several participants found it difficult to assess correctness due to unclear wording and insufficient contextual information.

\begin{customquote}
    \textit{“Maybe use of references [would improve correctness] to ‘back it up’ with requirements, standards, conventions, etc.”}
\end{customquote}

In the Date FUT, participants raised concerns about mismatches between explanations and boundary pairs. Specifically, one of the  explanations incorrectly described a change from three to four digits in the year field when none occurred (a hallucination), while others questioned the validity of negative years\footnote{Julia programming language allows for negative years.} (i.e., \textit{Before Christ} values) and found some explanations illogical.

\begin{customquote}
    \textit{``The explanation is not correct, the output field is 4 digits in both cases.”}
    
    \textit{“Date seems to be invalid. We cannot have a negative year.”}
\end{customquote}

\subsubsection{Completeness}

Participants generally found the explanations mostly complete but identified several gaps where additional detail or context was needed. Across FUTs, respondents frequently asked for clearer definitions of technical terms (e.g., in the Email FUT: “local part”, “RFC-5322”, “punycode”), explicit assumptions, and references to relevant standards or formulas (such as the BMI calculation). Many also requested broader example coverage --- especially for special cases, unrealistic values, and additional boundary pairs --- to better illustrate the reasoning behind each condition.

\begin{customquote}
    \textit{``Not only can it be hard to know if it is correct, it is also hard to know if this information is enough/sufficient for me to identify the boundaries. References and/or more information to justify the statement can be valuable."}
\end{customquote}

Participants also noted inconsistencies in the level of detail: some explanations covered only one side of a boundary, while others included unnecessary or confusing information (e.g., unrelated parameters or negative years in the Date FUT, i.e., \textit{Before Christ} in the Gregorian calendar). Overall, respondents emphasised that completeness requires not only accurate boundary descriptions but also sufficient context and justification to make them understandable and verifiable.

\begin{customquote}
    \textit{``Feels like in several situations the explanation can be more clear --- not only about the identified boundary but also about other related boundaries which are not mentioned."}
\end{customquote}

\subsubsection{Perceived Usefulness}

Participants highlighted the value of explanations in clarifying and contextualising boundary behaviours. However, they also identified several opportunities to enhance perceived usefulness.

For the Email FUT, respondents expressed that the explanations often added valuable insight, especially when highlighting differences not immediately evident from the boundary pairs alone. Yet, respondents mentioned that usefulness could improve with more explicit elaboration on valid versus invalid behaviours and removal of redundant information.

Regarding the Bytecount FUT, explanations were generally viewed as useful, although participants indicated that perceived usefulness was reduced when the boundary itself was unclear. Specifically, one boundary pair (input1 = -9999999999999999, input2 = -10000000000000000; output1 = 9999999999999999B, output2 = -10000000000000000B) was considered confusing. Its auto-generated explanation stated: \textit{“No boundary exists here; both inputs yield the same plain byte output.”} Participants criticised this explanation as not useful, but their feedback (quoted below) suggests their assessment was confounded by doubts about the boundary pair itself:

\begin{customquote}
\textit{“Q3.3 is not useful since it does not describe a boundary: both outputs belong to the same behavioral class.”}

\textit{“The explanation does not explain why this is being tested, since it doesn’t test a boundary.”}
\end{customquote}

For the Date FUT, participants pointed out cases where unclear or incorrect descriptions limited the practical value of the explanations. They suggested adding explicit reasoning behind observed formatting shifts in the FUT and including examples closer to real-world scenarios. 

Lastly, in the BMI FUT, perceived usefulness suffered when boundary pairs seemed unrealistic. Respondents called for clearer definitions of general boundary behaviors independent of specific input values.

\begin{customquote}
    \textit{``The examples were not realistic. You can't be 20 cm high but I understand that we are testing edge cases here."}
\end{customquote}

\subsubsection{Summary}
Mean scores were 3.7 out of 5 for clarity, completeness, and perceived usefulness, and 3.8 for correctness, indicating generally positive but improvable quality. 
Survey participants provided several suggestions to improve the current explanation format, emphasising the importance of clarity, structure, and contextual completeness. They recommended a more consistent structure across explanations, clearly indicating boundary values and incorporating concise examples for potential fixes. They also proposed adding external links to balance completeness with brevity and suggested extending explanations beyond single isolated boundaries to offer a broader perspective. Lastly, they suggested using simpler language to make explanations easier to understand.

Overall, respondents expressed interest in using a tool that provides auto-generated boundary explanations, emphasising benefits such as improved debugging capabilities, clearer context, and better understanding of system behaviours. They highlighted the potential for consistency with existing code and noted potential productivity gains through automated test generation and reduced debugging time. Some participants, however, remained cautious and expressed a preference for evaluating the tool’s practical usefulness themselves before fully trusting and adopting it.

\subsection{RQ2: Good Boundary Explanations}
\label{a:rq2}

\begin{figure}[H]
    \centering
    \includegraphics[width=\linewidth]{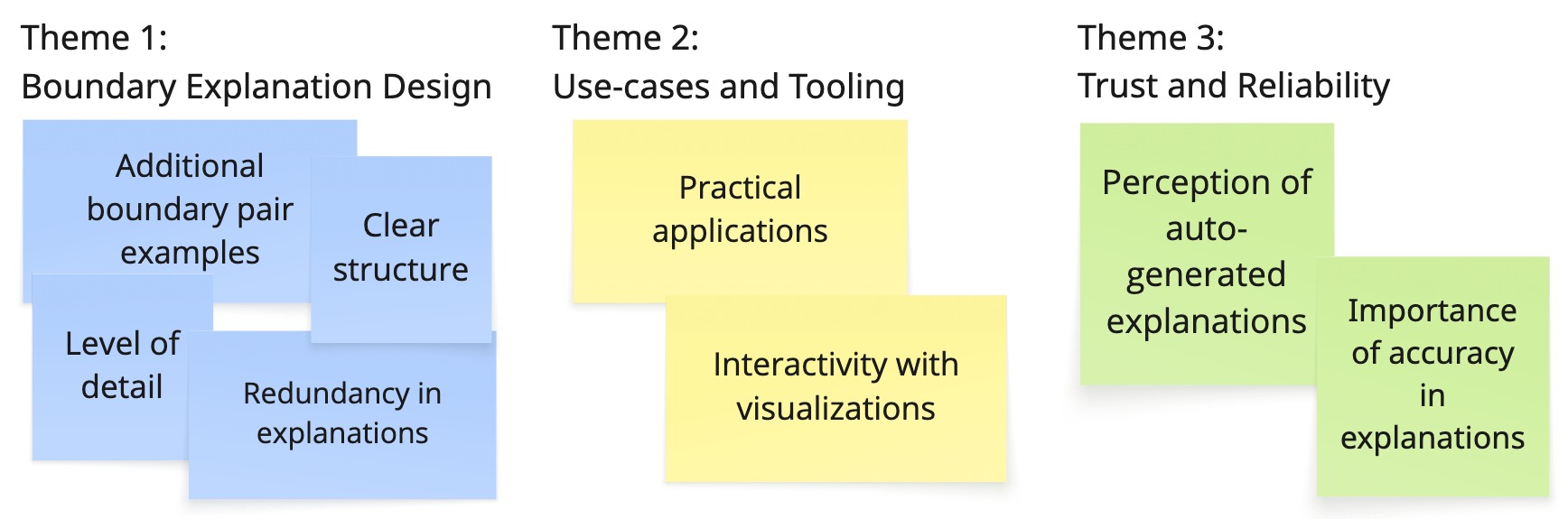}
    \caption{Themes and corresponding codes identified through thematic analysis of the interview data.}
    \label{fig:ta_codes}
\end{figure}

Based on the interviews, our thematic analysis revealed three interconnected themes that contain different aspects of boundary explanations (Figure \ref{fig:ta_codes}). The first theme centers on the elements constituting effective boundary explanations. The second theme addresses how these explanations can be practically integrated into real-world tasks and tools, thereby enhancing their applicability. Finally, participants’ perspectives on trust and reliability emerge as a third theme, highlighting that successful adoption of auto-generated explanations relies on their perceived accuracy. 

\subsubsection{Theme 1: Boundary Explanation Design}  
Participants emphasised that a well-designed boundary explanation goes beyond simply stating facts --- it should support understanding through diversified examples around the boundary, an appropriate level of detail, a clear focus, and an intuitive structure. Clearly stating what lies on either side of a boundary is important, as is specifying the expected behavior.

\begin{customquote}
    \textit{``[The explanation should state] what the boundaries are and the values on each side. I enjoyed the ones that were fairly structured and very clearly described."}
    \\ --P5
\end{customquote}

According to participants, a critical feature of effective boundary explanations is the inclusion of additional examples of boundary pairs close to the initially described boundary. Such examples would help illustrate how behaviors change across boundaries, making boundary behavior easier to grasp. 

\begin{customquote}
    \textit{``Having more explicit examples of values and how they trigger changes in output or expected output would be valuable."}
    \\ --P2
\end{customquote}

The level of detail also influenced perceived explanation quality. Participants noted that the appropriate amount of detail depends on the user’s experience and background knowledge. P1 highlighted the trade-off between readability and depth, suggesting a layered approach, while P2 supported the idea of expandable explanations that adapt to user needs. P6 added that more technically oriented users may prefer greater technical depth and links to authoritative specifications.

\begin{customquote}
   \textit{“There is a trade-off between explanation being brief enough to be readable, and the level of the details needed [for the user experience level]. Maybe a good suggestion for that would be having links that point to explanations of the terms.”}
   \\ --P1
\end{customquote}

Meanwhile, P3 noted that even concise explanations can be useful when they directly address the questions a user might have upon seeing a boundary case.

\begin{customquote}
    \textit{“When not obvious why something is boundary, then the explanation provides a rationale behind it. It’s basically when you look at the example the first things that come to your mind are questions. If the explanation is addressing that, then it would be useful.”} 
    \\ --P3
\end{customquote}

At the same time, participants warned against redundancy or irrelevant detail. Explanations that simply echoed what was already clear from the example were seen as unhelpful.

Finally, the structure of an explanation is seen as important for comprehension. Participants noted that how information is presented --- especially the sequencing of inputs, outputs, and transitions --- affects how intuitive the explanation feels.

\subsubsection{Theme 2: Use-cases and Tooling}  
Participants suggested some practical scenarios where boundary explanations could support everyday engineering tasks and broader goals like learning and documentation.
One idea was the integration of boundary explanations into developer workflows, such as debugging or test validation. Participants imagined tools that could automatically surface explanations during coding, continuous integration, or analysis, helping developers interpret unexpected behavior. For example, P2 suggested using them as part of a static analysis workflow, where they could offer immediate feedback during smoke testing. 

\begin{customquote}
   \textit{“They could run this as a static code analysis tool quickly and get a few of these explanations and then maybe could give hints to why there are strange behaviors”} 
   \\ --P2
\end{customquote}

P6 noted that explanations can be particularly valuable when failures occur and that test-related metadata should be surfaced alongside the boundary test case explanation.
   
\begin{customquote}
   \textit{“I think it would be quite useful, when a test fails, to know something about the test — for example, its running time, which is an important thing to note when explaining your tests.”}
   \\ --P6
\end{customquote}

Beyond tooling for active development, participants highlighted the value of boundary explanations for learning and documentation. For junior engineers or those unfamiliar with a system’s logic, the explanations were seen as helpful for understanding how inputs influence behavior. P3 and P4 expressed the idea of lightweight documentation that could clarify parameter constraints in function definitions or even generate useful assertion messages.

\begin{customquote}
    \textit{“[One can use boundary explanations for] documentation for methods parameters. [...] Maybe it would be useful to have it integrated into kind of plug-in."}
    \\ --P4
\end{customquote}

\begin{customquote}
\textit{“The use case for this kind of explanation is to serve as a supporting element to requirements or function descriptions.”}
\\ --[P5]
\end{customquote}

There was also interest in visualizing boundaries as a way to make explanations more intuitive when integrated in tools, particularly for training purposes. Participants noted that showing input spaces and where the boundary lies could deepen understanding. However, the participants cautioned that visuals should remain optional to avoid overwhelming experienced users.

\begin{customquote}
    \textit{“Visualizing the input space and where the boundary actually lies --- that would be great for learning."}
    \\ --P3
\end{customquote}

\subsubsection{Theme 3: Trust and Reliability in Auto-generated Explanations}  

Participants reflected on their trust in auto-generated boundary explanations, which was shaped by both the accuracy of the content and their own experience levels. Trust was not a fixed quality, but one that fluctuated based on how well explanations met expectations and avoided errors.

P1 noted that domain knowledge strongly influenced judgment: when less familiar with the topic, even simple clarifications could build confidence. Another participant observed that perceived clarity and completeness often depended on prior expertise.

\begin{customquote}
    \textit{“There were domain-specific errors that I wasn’t aware of and the explanations helped me a lot [...] looking at the boundary alone was not enough."}
    \\ --P1
\end{customquote}

Participants generally expected that the explanations were LLM-generated, and some recognized this early based on stylistic cues or recurring structure. This awareness did not, by itself, diminish trust, but it did shape expectations about potential limitations. Accuracy, however, proved essential: even small factual errors or hallucinations quickly undermined confidence. One hallucinated explanation led to markedly lower ratings for a boundary pair, underscoring how fragile trust can be. Repeated errors could easily erode confidence in LLM-based testing tools.

\begin{customquote}
    \textit{“Maybe then correctness would be one of the most important criteria [for trust]"}
    \\ --P4
\end{customquote}

Overall, participants were open to using auto-generated explanations, especially when these filled gaps in understanding. Yet their trust hinged on accuracy and relevance: inconsistent or hallucinated explanations, even if rare, were enough to raise doubts. Participants also cautioned against over-reliance, emphasizing that explanations should support rather than replace engineers’ reasoning. As a result, the explanations were viewed as useful but best used with critical attention.

\begin{customquote}
    \textit{``If we can get help identifying and explaining boundaries, it could, of course, be helpful, right? The risk in having explanations is that you might focus only on what is explained to you, and maybe you're not drawing your own conclusions, or analyzing it [software behavior] correctly."}
    \\ --P5
\end{customquote}

\section{Discussion}
\label{discussion}

LLMs can generate boundary explanations that software engineering practitioners and researchers consider moderately clear, correct, complete, and useful, indicating that LLM-based explanations hold promise for supporting automated test generation workflows.

To the best of our knowledge, no prior study has examined LLM–generated explanations for \emph{boundary value} testing. The nearest effort in the testing space is Kang~\textit{et~al.}~\cite{kang2025explainable}, who use an LLM-guided “scientific debugging’’ loop to explain failing tests; their participants appreciated concise wording but still questioned factual grounding. Beyond testing, evidence is limited: Sobania~\textit{et~al.} found that GPT-3.5 narrates both search-based and human patches with high surface accuracy yet thin rationale~\cite{sobania2023evaluating}, while d’Aloisio~\textit{et~al.} reported a similar clarity–completeness trade-off for quantum-algorithm summaries, even with added context~\cite{d2024exploring}. Our exploratory results echo these patterns: respondents judged boundary rationales easy to read and mostly correct, but asked for firmer justification, citations to authoritative sources, and extra examples that show how behavior changes across the edge. These signals hint that boundary explanations may demand adjustable depth and multi-example context before testers will rely on them.  Accordingly, we see this work as a first systematic step toward making input–output \emph{boundaries} transparent through LLM explanations.

Survey respondents rated auto‐generated explanations between 3.5 and 4.0 on a five‐point Likert scale across clarity, correctness, completeness, and perceived usefulness, despite minimal prompt engineering. While this indicates moderate quality overall, the distribution shows more agreement than disagreement (63.5\% of ratings were 4–5 versus 17\% rated 1–2), but the relatively high standard deviations (1.1–1.3) reveal substantial variance in perceptions. Interview participants confirmed the appropriateness of the four evaluation criteria, suggesting that our evaluation framework captures essential qualities of effective boundary explanations.

The positive reception encourages further development, especially since participants' concerns about correctness targeted the boundary examples themselves rather than the explanations and that the LLM sometimes also commented on the suitability of the underlying example. One survey respondent argued that a boundary should be treated as a \emph{set of values} and cautioned against conflating valid/invalid outputs with the boundary itself. The same respondent noted that boundaries depend on underlying implementation (no code was provided). These comments also illustrate that, in some responses, attention shifted from evaluating the explanations to disputing the quality of the boundary pairs.

This suggests great potential for further improvement through targeted prompt refinements, a refined study design, and more sophisticated LLM-based systems that not only explain pre-selected boundaries but can also be used to find and\slash or select them.

LLM‐based boundary explanations can bridge gaps in domain knowledge by providing immediate, contextualized rationales for why certain input pairs represent boundaries. Practically, integrating such explanations into static analysis or test validation tools could help developers identify unexpected behaviors more efficiently, reducing debugging time and improving test coverage. From a research perspective, our findings clarify requirements for future automated explanation systems, laying a foundation for evaluating and improving LLM‐driven test explanation tools. 
Overall, our results also advance our understanding of human–LLM collaboration in software testing.

An alternative interpretation of our results is that participants, many of whom were not routinely exposed to automated boundary value testing, may have rated explanations favorably because a tool had succeeded in identifying boundary pairs, irrespective of explanation quality. In other words, initial impressions of usefulness might reflect appreciation for automated boundary detection rather than for the explanatory content itself. Although we tried to mitigate this threat, and discuss it in more depth in the threats to validity section below, it suggests that future studies should control for familiarity with boundary testing tools to better isolate judgments of explanation quality from judgments of boundary identification.

Open-ended survey and interview feedback revealed clear opportunities for improving auto-generated explanations, many of which are addressable via prompt engineering or multi-agent LLM setups. Suggestions included defining technical terms, structuring explanations from valid to invalid behavior, and linking to external standards to enhance completeness and trust. These align with known strengths of prompt engineering and LLM integration with external knowledge sources~\cite{white2023prompt}. However, more complex issues --- such as explaining intricate error formats or avoiding redundant statements --- may require deeper, domain-specific reasoning beyond prompt-level adjustments.

\subsection{Requirements for LLM-based Boundary Explanation Tools}

Drawing on survey feedback and thematic analysis of interview data, we propose the following set of requirements for LLM-based boundary explanation tools. These requirements can be viewed as a starting point for both tool designers and researchers in the automated software engineering community:

\begin{itemize}
\item \textbf{Adaptive Detail Levels:} Explanations should adjust verbosity based on user expertise and use case. Provide concise summaries for experienced users, while offering expandable sections (e.g., via hyperlinks or collapsible text) that define technical terms and elaborate on domain‐specific concepts for less experienced users.

\item \textbf{Authoritative References:} Whenever possible, link technical terminology (e.g., zero‐padding, RFC‐5322, or BMI formulas) directly to standards, documentation pages, or established specifications. This builds trust and allows users to verify boundary logic independently. A testing tool could also link to relevant pieces of source code to invite direct review of the implementation that leads to the behaviour.

\item \textbf{Explicit Structure:} Present information in a predictable sequence: (1) state the valid boundary condition, (2) state the invalid boundary condition, (3) show input and output pairs, (4) provide justification reasoning, and (5) link to relevant documentation or code. Consistency in structure reduces cognitive load and makes it easier to compare multiple boundaries.

\item \textbf{Multiple Examples Around the Boundary:} Include at least one additional boundary pair on either ``side'' of the primary boundary pair. Demonstrating how behaviour changes incrementally helps users form an intuitive mental model of the input space. This indicates that explanations might benefit by clustering and selecting multiple boundary pairs rather than relying on only a single one.

\item \textbf{Contextual Justification:} Whenever a boundary depends on domain rules (e.g., negative years in Gregorian calendars, out‐of‐bounds byte counts), explanations should briefly describe the underlying rationale or assumptions (``of'' the LLM). This can help avoid hallucinations (e.g., misstating digit lengths) and clarify why the model thinks a boundary exists. Potentially, the user should then be allowed to ``lift'' the assumption and re-generate the explanation.

\item \textbf{Interactive, “Live” Conversations:} Rather than providing static text, test explanation systems should allow users to query or challenge specific explanation elements in a dialogue with the LLM. Live interaction enables users to request further clarification (e.g., “Why is a year $< 0$ considered invalid?”) or to resolve apparent contradictions on demand.

\item \textbf{Integration with Tooling Workflows:} Explanations should be surfaced in situ, such as IDE plugins, static analysis dashboards, or test generation GUIs, so that users can access boundary rationales without context switching. Optional visualizations (e.g., input‐space plots) can further enhance comprehension, especially for training or documentation purposes, but also in general for exploring boundaries~\cite{dobslaw2020boundary}.
    
\end{itemize}

These requirements are not exhaustive but represent a first step toward defining a concrete set of expectations for future LLM‐based boundary explanation systems. In particular, the emphasis on adaptive detail levels and interactive dialogues aligns with emerging trends in automated SE research on human–AI collaboration and situational AI support in developer workflows \cite{khojah2024beyond}.

\subsubsection{Preliminary Test of Requirements via Prompting}
To explore the practical utility of the proposed requirements, we minimally modified the original prompt --- removing the character limit and incorporating some of the requirements --- while keeping the same model and temperature. This follow-up test yielded improved explanations. For instance, in response to earlier feedback from participants about an unclear error in the \textbf{Bytecount} FUT, the LLM generated the following boundary explanation:

\begin{customquote}
Boundary: The transition occurs at bytes = 999...40.

I/O pairs:

$bytes = 999...38 \rightarrow 999.9 EB$

$bytes = 999...39 \rightarrow 1000.0 EB$

$bytes = 999...40 \rightarrow BoundsError(kMGTPE, 7)$

$bytes = 999...41 \rightarrow BoundsError(kMGTPE, 7)$

Assumption: The function supports up to exabytes (EB), using the unit list "kMGTPE" (kilobyte to exabyte), and does not handle values requiring a higher unit.

Crossing this threshold (from 999...39 to 999...40) causes the function to attempt to access a unit beyond exabyte, resulting in a bounds error. This is because 1000 exabytes is not representable with the available units, so the function fails when the input exceeds the maximum representable value in exabytes.
\end{customquote}

The revised explanation includes an additional boundary pair, clearly states assumptions and follows a structured format --- features aligned with our proposed requirements. While this is only a preliminary trial, it suggests that advanced prompting can improve boundary explanations and warrants further investigation. Additional examples are available in our replication package.

While our study focused specifically on boundary testing, several of the presented requirements --- such as structured explanations, incorporation of authoritative references, and interactive dialogues --- are likely valuable in explaining other types of test cases (e.g., equivalence classes, mutation testing) or even debugging outcomes (e.g., stack‐trace interpretation). We suggest that researchers investigate how these requirements transfer to broader software testing and verification scenarios. However, further empirical work is needed to confirm generality beyond boundary explanations.

\subsection*{Threats to Validity}
We discuss threats to validity using standard categories to provide a balanced assessment of our findings.

\subsubsection*{Construct Validity}

Our evaluation used four criteria to assess explanation quality: clarity, correctness, completeness, and usefulness. All four dimensions were assessed through subjective human ratings, which may introduce interpretation bias and variance across participants. Participants also noted that these dimensions interact: low clarity, for instance, can reduce perceived usefulness. We addressed this interdependence by analyzing each criterion separately and triangulating results with qualitative feedback, though some overlap likely persists.

Explanation quality also depends on the selected boundary pairs. All authors independently verified that the 20 examples represent diverse types (VV, VE, EE), but different or more complex boundaries may expose additional challenges not captured here.

We used a single general-purpose model (GPT-4.1) with minimal prompt engineering, aiming to test whether LLMs can \textit{in principle} generate meaningful boundary explanations. Therefore, our findings should be viewed as indicative rather than generalizable to all LLMs. Testing multiple models or systematically varying prompts would have required significantly more participants or unrealistically many questions per participant. To support our choice, we compared GPT-4.0 and 4.1 outputs via semantic similarity and found them highly comparable under the same prompt. Still, specialized models or alternative prompting may yield different results and warrant systematic exploration in future work.

We constrained explanations to 150 characters to reduce cognitive load across 20 items and four criteria. This practical limit helped with evaluation but introduced trade-offs: our follow-up trial with longer outputs showed that more structured and informative explanations may require additional space. Future work should explore adaptive verbosity and advanced prompting techniques, as even minimal prompt modifications yielded clearer and more complete explanations.

\subsubsection*{Internal Validity}

Some participants may have rated explanations poorly due to issues with the boundary examples themselves. Presenting each boundary without its explanation in a pilot confirmed that the pairs were understandable, and the survey always showed the pair and explanation together to keep context constant. However, we acknowledge a potential confound: unclear or questionable boundary cases may have negatively influenced perceived explanation quality. Moreover, some positive ratings might reflect appreciation for boundary identification rather than explanation quality --- for instance, if participants were impressed that a system could identify boundaries at all. Future studies could address this by separating boundary evaluation from explanation evaluation --- e.g., by pre-validating boundary pairs with independent raters or presenting multiple explanations for the same boundary to isolate effects.

For the Email FUT, some boundary pairs were initially suggested by an LLM, introducing a potential self-referential bias. To mitigate this, all boundary pairs were independently reviewed by the authors to ensure semantic validity and plausibility.

Social desirability bias presents another threat: participants may overstate the value of novel (AI) tools. We mitigated this by keeping surveys anonymous and revealing the LLM's provenance only during interviews, though we cannot eliminate this effect entirely.

For interviews, a single researcher conducted the initial coding. To reduce personal bias, the full author team reviewed all codes and themes until reaching consensus.

\subsubsection*{External Validity}
Our convenience sample (27 respondents, 6 interviewees), although balanced between industry and academia, is not representative of all software roles or domains. Replicating the study with larger, stratified samples is needed. We considered only four small FUTs implemented in Julia. Our use of simple black-box functions may not capture the challenges present in real-world codebases, such as error handling, program state, and concurrency. Moreover, we studied only textual explanations. Visual or interactive modes — suggested by interviewees — could change perceived usefulness and should be explored further.

\subsubsection*{Conclusion Validity}
We provide a complete replication package, including prompts, boundary pairs, anonymised raw data, and analysis scripts. This enables independent re-analysis and verification of our findings.

Two factors threaten long-term repeatability. First, our reliance on a single LLM version poses challenges because commercial models evolve over time. Future researchers may obtain different results with updated model versions.

Second, LLM non-determinism could affect explanation stability. We did not measure this stability empirically, leaving it as an explicit topic for future work.

Overall, our mitigation strategies reduce several validity threats, but the remaining risks constrain our conclusions. These limitations point to clear directions for future research: measuring explanation stability, evaluating alternative models with more advanced prompting, and studying broader participant populations.

\section{Conclusion}
\label{conclusions}
In this paper, we studied LLM-generated explanations of software boundaries and what characterizes good ones that create value for software professionals. Using a sequential mixed-methods design with a survey and interviews analyzed through thematic analysis, we found:
\begin{itemize}
    \item LLM-generated explanations were generally perceived as useful and show potential for integration into testing tools.
    \item The four categories, clarity, correctness, completeness, and perceived usefulness, were overall assessed as effective criteria for evaluating explanation quality.
    \item We distilled 7-item requirements that define concrete design principles for future LLM-based boundary explanation tools, offering enduring value beyond the specific prompts or models evaluated here.
\end{itemize}

These findings indicate that LLM-generated explanations could enhance testing workflows, particularly by improving test case selection and implementation efficiency. However, generalizability cannot yet be assumed; further studies are needed to validate and refine what constitutes effective boundary explanations and how they can be best produced or applied in practice. Future work should focus on improved prompting, broader practitioner feedback, and turning the derived requirements into actionable guidelines for practical tools. Finally, advancing trustworthy AI-supported testing tools will require multidisciplinary input, including HCI expertise, to address issues of usability, trust, and reliability.

\begin{acks} 
This work was supported by the Wallenberg AI, Autonomous Systems and Software Program (WASP) funded by the Knut and Alice Wallenberg Foundation. Robert Feldt and Felix Dobslaw have also been supported by the Swedish Scientific Council (No. 2020-05272, `Automated boundary testing for Quality of AI/ML models'). ChatGPT and Claude.ai was utilized to improve the phrasing of some parts of the text, originally written by the authors.
\end{acks}

\balance
\bibliographystyle{ACM-Reference-Format}
\bibliography{references}

\end{document}